\documentclass[reprint,showpacs,preprintnumbers,nofootinbib,amsmath,amssymb,aps,prd,floatfix]{revtex4-1}
\pdfoutput=1
\usepackage{graphicx}
\usepackage{bm}
\usepackage{subfigure}
\usepackage{url}
\usepackage[hyperindex]{hyperref}
\usepackage{color}
\usepackage[ddmmyy,24hr]{datetime}
\usepackage{bigdelim}
\usepackage{booktabs}
\usepackage{dcolumn}
\usepackage{multirow}
\usepackage{subfigure}
\usepackage{cancel}
\usepackage{stackrel}
\usepackage{paralist}
\usepackage{xspace}
\newcommand{\nua}[1]{\ensuremath{\rlap{\kern-2.5pt\ensuremath{\overset{\scriptscriptstyle(-)}{\phantom{\nu}}}}{\ensuremath{{\nu}_{#1}}}}}

\usepackage{enumitem}

\makeatletter
\def\namedlabel#1#2{\begingroup
    #2%
    \def\@currentlabel{#2}%
    \phantomsection\label{#1}\endgroup
}
\newcommand{\nucleus}{\mathcal{N}}
\makeatother
\begin{document}

\title{Average CsI neutron density distribution from COHERENT data}

\author{M. Cadeddu}
\affiliation{Dipartimento di Fisica, Universit\`{a} degli Studi di Cagliari,
and
INFN, Sezione di Cagliari,
Complesso Universitario di Monserrato - S.P. per Sestu Km 0.700,
09042 Monserrato (Cagliari), Italy}

\author{C. Giunti}
\affiliation{INFN, Sezione di Torino, Via P. Giuria 1, I--10125 Torino, Italy}

\author{Y.F. Li and Y.Y. Zhang}
\affiliation{Institute of High Energy Physics,
Chinese Academy of Sciences,
and
School of Physical Sciences, University of Chinese Academy of Sciences, Beijing 100049, China}

\date{23 January 2018}

\begin{abstract}
Using the coherent elastic neutrino-nucleus scattering data
of the COHERENT experiment,
we determine for the first time the average neutron rms radius
of $^{133}\text{Cs}$ and $^{127}\text{I}$.
We obtain the practically model-independent value
$
R_{n}
=
5.5
{}^{+0.9}_{-1.1}
\,
\text{fm}
$
using the symmetrized Fermi and Helm form factors.
We also point out that the COHERENT data show a
$2.3\sigma$
evidence of the nuclear structure suppression of the full coherence.
\end{abstract}


\maketitle

The COHERENT experiment \cite{Akimov:2017ade}
observed for the first time coherent elastic neutrino-nucleus scattering
with a small scintillator detector made of sodium-doped CsI
exposed to a low-energy neutrino flux generated in the
Spallation Neutron Source at Oak Ridge National Laboratory.
Coherent elastic neutrino-nucleus scattering can occur if
$q R \ll 1$,
where $q=|\vec{q}|$ is the three-momentum transfer
and $R$ is the nuclear radius~\cite{Freedman:1973yd,Freedman:1977xn}.

The coherent elastic scattering of a neutrino with a nucleus can be observed
by measuring very low values of the nuclear kinetic recoil energy $T$.
For $T \ll E$,
where $E$ is the neutrino energy,
we have
$q^2 \simeq 2 M T$,
where $M$ is the nuclear mass,
and $T_{\text{max}} \simeq 2 E^2 / M$
\cite{Drukier:1983gj}.
For a nucleus with mass
$M \approx 100 \, \text{GeV}$
and radius
$R \approx 5 \, \text{fm}$,
elastic neutrino-nucleus scattering is coherent for
$T \ll ( 2 M R^2 )^{-1} \approx 10 \, \text{keV}$
and it is required to have a neutrino beam
with energy of the order of
$\sqrt{M T / 2} \approx 20 \, \text{MeV}$.

The differential cross section for coherent elastic scattering
of a neutrino with a nucleus $\nucleus$ with $Z$ protons and $N$ neutrons
is given by~\cite{Drukier:1983gj,Barranco:2005yy,Patton:2012jr,Papoulias:2015vxa}
\begin{align}
\dfrac{d\sigma_{\nu\text{-}\nucleus}}{d T}(E,T)
\simeq
\null & \null
\dfrac{G_{\text{F}}^2 M}
{4 \pi}
\left(
1 - \dfrac{M T}{2 E^2}
\right)
\nonumber
\\
\null & \null
\times
\left[
N
F_{N}(q^2)
-
\epsilon
Z
F_{Z}(q^2)
\right]^2
,
\label{cs01}
\end{align}
where
$G_{\text{F}}$ is the Fermi constant,
$M$ is the nuclear mass,
$F_{N}(q^2)$ and $F_{Z}(q^2)$
are, respectively,
the nuclear neutron and proton form factors,
and
$
\epsilon
=
1 - 4 \sin^2 \vartheta_{\text{W}}
=
0.0454 \pm 0.0003
$,
using the low-energy PDG value of the weak mixing angle $\vartheta_{\text{W}}$
\cite{PDG-2016}.
Because of the small value of $\epsilon$,
the neutron contribution is dominant.
Hence, measurements of the process give information on the nuclear neutron form factor,
which is more difficult to obtain than the information on the proton
nuclear form factor,
that can be obtained with elastic electron-nucleus scattering
and other electromagnetic processes
(see Refs.~\cite{Fricke:1995zz,Angeli:2013epw}).
Knowledge of these form factors is important,
because form factors are the Fourier transform of the corresponding
charge distribution.
Electromagnetic processes probe the nuclear proton distribution,
whereas neutral-current weak interaction processes
are mainly sensitive to the nuclear neutron distribution.
Also hadron scattering experiments give information on the nuclear neutron distribution,
but their interpretation depends on the model used to describe non-perturbative strong interactions
(see Refs.~\cite{GarciaRecio:1991wk,Starodubsky:1994xt,Trzcinska:2001sy,Clark:2002se}).
Before the COHERENT experiment,
the only measurement of the nuclear neutron distribution
with neutral-current weak interactions was done with parity-violating
electron scattering on $^{\text{208}}\text{Pb}$
in the PREX experiment~\cite{Abrahamyan:2012gp}.

\begin{figure*}[!t]
\centering
\includegraphics*[width=\linewidth]{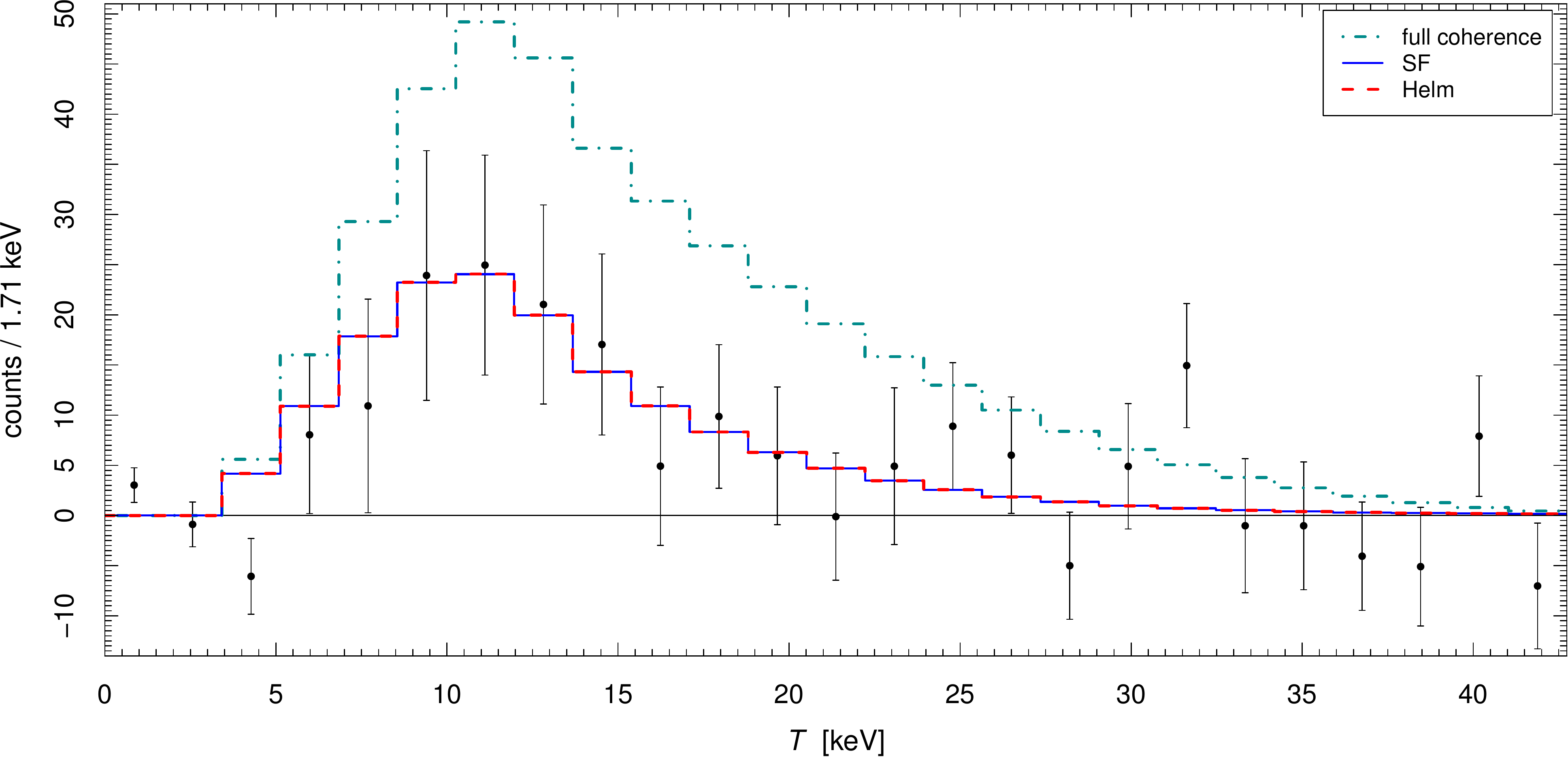}
\caption{ \label{fig:data}
COHERENT data \cite{Akimov:2017ade} versus the nuclear kinetic recoil energy $T$.
The histograms represent the theoretical prediction
in the case of full coherence (cyan dash-dotted)
and the best fits obtained using the symmetrized Fermi (SF) distribution (blue solid)
and Helm (red dashed) form factors.
}
\end{figure*}

The measurement of the nuclear neutron density distribution
is a topic of broad interest in the physics community.
In particular, the corresponding
rms radius $R_n$
and the difference between $R_n$ and the rms radius $R_p$ of the proton distribution
(the so-called ``neutron skin'')
are crucial ingredients of the nuclear matter Equation of State (EOS), which plays an essential role in understanding several processes, like nuclei in laboratory experiments, heavy ion collisions, and the structure and evolution of compact astrophysical objects as neutron stars
(see Refs.~\cite{Brown:2000pd,Horowitz:2000xj,Reinhard:2010wz,Tsang:2012se,Hagen:2015yea}).

In the case of the COHERENT experiment,
the coherent elastic scattering is measured on
$^{133}\text{Cs}$
and
$^{127}\text{I}$,
which contribute incoherently,
leading to the total cross section
\begin{equation}
\dfrac{d\sigma_{\nu\text{-}\text{CsI}}}{d T}
=
\dfrac{d\sigma_{\nu\text{-}\text{Cs}}}{d T}
+
\dfrac{d\sigma_{\nu\text{-}\text{I}}}{d T}
,
\label{cs02}
\end{equation}
with
$N_{\text{Cs}} = 78$,
$Z_{\text{Cs}} = 55$,
$N_{\text{I}} = 74$, and
$Z_{\text{I}} = 53$.
We neglect the small axial contribution due to the unpaired valence proton~\cite{Barranco:2005yy}.

The proton and neutron form factors
are the Fourier transform of the nuclear proton and neutron densities.
The proton structures of $^{133}\text{Cs}$ and $^{127}\text{I}$ have been studied with
muonic atom spectroscopy~\cite{Fricke:1995zz}
and the data were fitted with Fermi density distributions of the form
\begin{equation}
\rho_{\text{F}}(r)
=
\dfrac{ \rho_{0} }{ 1 + e^{ ( r - c ) / a } }
,
\label{fermi}
\end{equation}
where
$\rho_{0}$ is a normalization factor
and
$a$ is a parameter which quantifies the surface thickness
$t = 4 a \ln 3$,
which was fixed at 2.30 fm.
The fit of the data yielded
$c_{\text{Cs}} = 5.6710 \pm 0.0001 \, \text{fm}$
and
$c_{\text{I}} = 5.5931 \pm 0.0001 \, \text{fm}$,
which correspond to the proton rms radii
\begin{align}
\null & \null
R_{p}^{\text{Cs}} = \langle r_{p}^2 \rangle_{\text{Cs}}^{1/2} = 4.804 \, \text{fm}
,
\label{RpCs}
\\
\null & \null
R_{p}^{\text{I}} = \langle r_{p}^2 \rangle_{\text{I}}^{1/2} = 4.749 \, \text{fm}
.
\label{RpI}
\end{align}
Hence,
the proton structures of $^{133}\text{Cs}$ and $^{127}\text{I}$ are similar.
Since we expect that also their neutron structures are similar
and the current uncertainties of the COHERENT data do not allow to distinguish between them,
we consider in Eq.~(\ref{cs02}) the approximation
\begin{equation}
F_{N,\text{Cs}}(q^2)
\simeq
F_{N,\text{I}}(q^2)
\simeq
F_{N}(q^2)
.
\label{ffn1}
\end{equation}
We fitted the COHERENT data under this approximation assuming proton form factors
$F_{Z}(q^2)$
for $^{133}\text{Cs}$ and $^{127}\text{I}$
given by the Fourier transform of a symmetrized Fermi (SF) distribution
$\rho_{\text{SF}}(r) = \rho_{\text{F}}(r) + \rho_{\text{F}}(-r) - 1$,
which is practically equivalent to a Fermi distribution
and gives an analytic expression for the form factor~\cite{Piekarewicz:2016vbn}:
\begin{align}
F_{Z}^{\text{SF}}(q^2)
=
\null & \null
\dfrac{ 3 }{ q c \left[ ( q c )^2 + ( \pi q a )^2 \right] }
\left[
\dfrac{ \pi q a }{ \sinh( \pi q a ) }
\right]
\nonumber
\\
\null & \null
\times
\left[
\dfrac{ \pi q a \sin( q c ) }{ \tanh( \pi q a ) }
-
q c \cos( q c )
\right]
.
\label{ffzSF}
\end{align}

In order to get information on the neutron distribution of $^{133}\text{Cs}$ and $^{127}\text{I}$
in the approximation in Eq.~(\ref{ffn1}),
we considered the following parameterizations of the neutron form factor
$F_{N}(q^2)$:

\begin{enumerate}

\item
A symmetrized Fermi form factor
$F_{N}^{\text{SF}}(q^2)$
analogous to that in Eq.~(\ref{ffzSF}).
In this case,
the neutron rms radius is given by
\begin{equation}
R_{n}^2 = \dfrac{3}{5} \, c^2 + \dfrac{7}{5} \, ( \pi a )^2
.
\label{RnSF}
\end{equation}
Since the COHERENT data are not sensitive to the surface thickness,
we consider the same value of
$t = 2.30 \, \text{fm}$
as for the proton form factor.
We verified that the results of the fit are practically independent of
small variations of the value of the surface thickness.

\item
The Helm form factor~\cite{Helm:1956zz}
\begin{equation}
F_{N}^{\text{Helm}}(q^2)
=
3
\,
\dfrac{j_{1}(q R_{0})}{q R_{0}}
\,
e^{- q^2 s^2 / 2}
,
\label{ffnHelm}
\end{equation}
where
$
j_{1}(x) = \sin(x) / x^2 - \cos(x) / x
$
is the spherical Bessel function of order one
and $R_{0}$ is the box (or diffraction) radius.
In this case,
the neutron rms radius is given by
\begin{equation}
R_{n}^2 = \dfrac{3}{5} \, R_{0}^2 + 3 s^2
.
\label{RnHelm}
\end{equation}
The parameter $s$ quantifies the surface thickness.
In this case we consider the value $s = 0.9 \, \text{fm}$
which was determined for the proton form factor of similar nuclei
\cite{Friedrich:1982esq}.
Also in this case,
we verified that the results of the fit are practically independent of
small variations of the value of the surface thickness.

\end{enumerate}

We fitted the COHERENT data in Fig.~3A of Ref.~\cite{Akimov:2017ade}
with the least-squares function
\begin{align}
\chi^2
=
\null & \null
\sum_{i=4}^{15}
\left(
\dfrac{
N_{i}^{\text{exp}}
-
\left(1+\alpha\right) N_{i}^{\text{th}}
-
\left(1+\beta\right) B_{i}
}{ \sigma_{i} }
\right)^2
\nonumber
\\
\null & \null
+
\left( \dfrac{\alpha}{\sigma_{\alpha}} \right)^2
+
\left( \dfrac{\beta}{\sigma_{\beta}} \right)^2
.
\label{chi}
\end{align}
For each energy bin $i$,
$N_{i}^{\text{exp}}$
and
$N_{i}^{\text{th}}$
are, respectively,
the experimental and theoretical number of events,
$B_{i}$ is the estimated number of background events
extracted from Fig.~S13 of Ref.~\cite{Akimov:2017ade}, and
$\sigma_{i}$ is the statistical uncertainty.
$\alpha$ and $\beta$
are nuisance parameters which quantify,
respectively,
the systematic uncertainty of the signal rate
and
the systematic uncertainty of the background rate.
The corresponding standard deviations
are
$\sigma_{\alpha} = 0.28$
and
$\sigma_{\beta} = 0.25$
\cite{Akimov:2017ade}.
We did not considered the first three energy bins
in Fig.~3A of Ref.~\cite{Akimov:2017ade},
which do not give any information on
neutrino-nucleus scattering
because they correspond to the detection of less than 6 photoelectrons,
for which the acceptance function in Fig.~S9 of Ref.~\cite{Akimov:2017ade}
vanishes.
We considered only the 12 energy bins from $i=4$ to $i=15$
for which the COHERENT collaboration fitted the quenching factor
in Fig.~S10 of Ref.~\cite{Akimov:2017ade}
and obtained the linear relation between the observed number of photoelectrons $N_{\text{PE}}$
and the nuclear kinetic recoil energy $T$ given by
\begin{equation}
N_{\text{PE}}
=
1.17
\left( \dfrac{T}{\text{keV}} \right)
.
\label{PE}
\end{equation}

The theoretical number of coherent elastic scattering events
$N_{i}^{\text{th}}$
in each energy bin $i$
depends on the nuclear neutron form factor
and it is given by
\begin{equation}
N_{i}^{\text{th}}
=
N_{\text{CsI}}
\int_{T_{i}}^{T_{i+1}} d T
\int_{E_{\text{min}}} d E
\,
A(T)
\,
\frac{d N_{\nu}}{d E}
\,
\dfrac{d\sigma_{\nu\text{-}\text{CsI}}}{d T}
.
\label{Nth}
\end{equation}
where $N_{\text{CsI}}$
is the number of CsI
in the detector
(given by
$ N_{\text{A}} M_{\text{det}} / M_{\text{CsI}}$,
where
$ N_{\text{A}} $
is the Avogadro number,
$ M_{\text{det}} = 14.6 \,\text{kg} $,
is the detector mass, and
$ M_{\text{CsI}} = 259.8 $
is the molar mass of CsI),
$ E_{\text{min}} = \sqrt{M T / 2} $,
$A(T)$
is the acceptance function given in Fig.~S9 of Ref.~\cite{Akimov:2017ade}
and
$d N_{\nu} / d E$
is the neutrino flux integrated over the experiment lifetime.
Neutrinos at the Spallation Neutron Source consist of a prompt component of monochromatic
$\nu_\mu$ from stopped pion decays,
$\pi^+\to \mu^++\nu_\mu$,
and two delayed components of
$\bar{\nu}_\mu$ and $\nu_e$
from the subsequent muon decays, $\mu^+\to e^+ + \bar{\nu}_\mu + \nu_e$.
The total flux $d N_{\nu} / d E$ is the sum of
\begin{align}
\frac{d N_{\nu_{\mu}}}{d E}
=
\null & \null
\eta
\,
\delta\!\left(
E - \dfrac{ m_{\pi}^2 - m_{\mu}^2 }{ 2 m_{\pi} }
\right)
,
\label{numu}
\\
\frac{d N_{\nu_{\bar\mu}}}{d E}
=
\null & \null
\eta
\,
\dfrac{ 64 E^2 }{ m_{\mu}^3 }
\left(
\dfrac{3}{4} - \dfrac{E}{m_{\mu}}
\right)
,
\label{numubar}
\\
\frac{d N_{\nu_{e}}}{d E}
=
\null & \null
\eta
\,
\dfrac{ 192 E^2 }{ m_{\mu}^3 }
\left(
\dfrac{1}{2} - \dfrac{E}{m_{\mu}}
\right)
,
\label{nue}
\end{align}
for
$E \leq m_{\mu} / 2 \simeq 52.8 \, \text{MeV}$,
with the normalization factor
$ \eta = r N_{\text{POT}} / 4 \pi L^2 $,
where
$r=0.08$ is the number of neutrinos per flavor
that are produced for each proton on target,
$ N_{\text{POT}} = 1.76\times 10^{23} $
is the number of proton on target
and $ L = 19.3 \, \text{m} $
is the distance between the source and the COHERENT CsI detector~\cite{Akimov:2017ade}.

\begin{figure}[!t]
\centering
\includegraphics*[width=\linewidth]{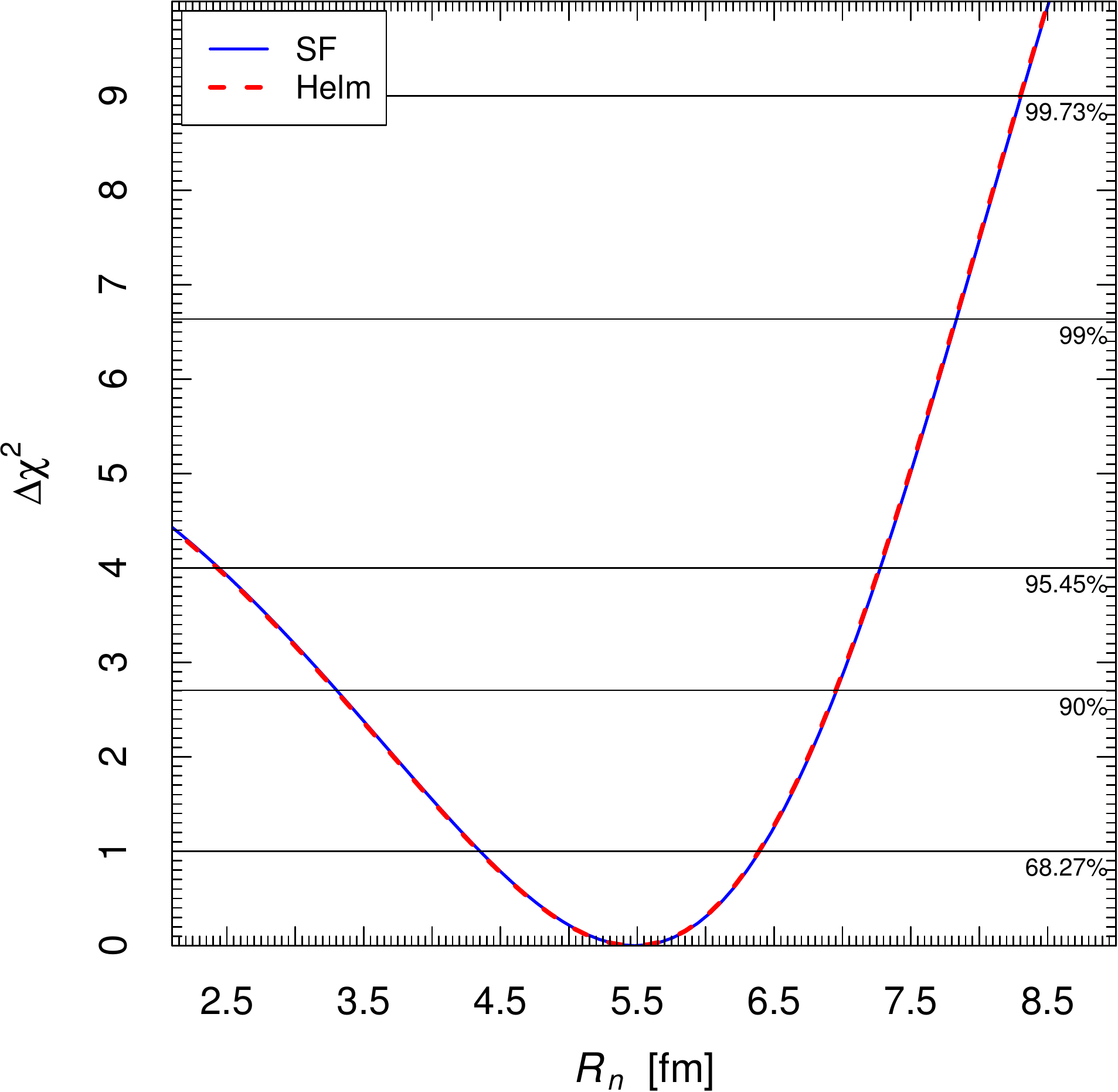}
\caption{ \label{fig:rn-chi}
$\Delta\chi^2 = \chi^2 - \chi^2_{\text{min}}$
as a function of the neutron rms radius $R_{n}$
obtained from the fit of the data of the COHERENT experiment \cite{Akimov:2017ade}
using the symmetrized Fermi (SF) and Helm form factors.
}
\end{figure}

Figure~\ref{fig:data}
shows the COHERENT data as a function of the nuclear kinetic recoil energy $T$.
We first compared the data with the predictions in the case of full coherence,
i.e. all nuclear form factors equal to unity.
Figure~\ref{fig:data} shows that the corresponding histogram
does not fit the data.
Hence,
albeit the COHERENT data represent the first measurement of
coherent elastic neutrino-nucleus scattering,
the scattering is not fully coherent and the data give information
on the nuclear structure.
Indeed, the COHERENT collaboration \cite{Akimov:2017ade} explained the data
using the form factor in Ref.~\cite{Klein:1999qj} with fixed value of
the parameters,
i.e. assuming the value of the nuclear rms radius.

\begin{table*}[t!]
\centering
\begin{tabular}{l|ccc|ccc|ccc}
&
\multicolumn{3}{c|}{$^{133}\text{Cs}$}
&
\multicolumn{3}{c}{$^{127}\text{I}$}
&
\multicolumn{3}{c}{CsI}
\\
Model
&
$R_{p}$
&
$R_{n}$
&
$R_{n}-R_{p}$
&
$R_{p}$
&
$R_{n}$
&
$R_{n}-R_{p}$
&
$R_{p}$
&
$R_{n}$
&
$R_{n}-R_{p}$
\\
\hline
SHF SkM* \cite{Bartel:1982ed}
&
4.76
&
4.90
&
0.13
&
4.71
&
4.84
&
0.13
&
4.73
&
4.86
&
0.13
\\
SHF SkP \cite{Dobaczewski:1983zc}
&
4.79
&
4.91
&
0.12
&
4.72
&
4.84
&
0.12
&
4.75
&
4.87
&
0.12
\\
SHF SkI4 \cite{Reinhard:1995zz}
&
4.73
&
4.88
&
0.15
&
4.67
&
4.81
&
0.14
&
4.70
&
4.83
&
0.14
\\
SHF Sly4 \cite{Chabanat:1997un}
&
4.78
&
4.90
&
0.13
&
4.71
&
4.84
&
0.13
&
4.73
&
4.87
&
0.13
\\
SHF UNEDF1 \cite{Kortelainen:2011ft}
&
4.76
&
4.90
&
0.15
&
4.68
&
4.83
&
0.15
&
4.71
&
4.87
&
0.15
\\
RMF NL-SH \cite{Sharma:1993it}
&
4.74
&
4.93
&
0.19
&
4.68
&
4.86
&
0.19
&
4.71
&
4.89
&
0.18
\\
RMF NL3 \cite{Lalazissis:1996rd}
&
4.75
&
4.95
&
0.21
&
4.69
&
4.89
&
0.20
&
4.72
&
4.92
&
0.20
\\
RMF NL-Z2 \cite{Bender:1999yt}
&
4.79
&
5.01
&
0.22
&
4.73
&
4.94
&
0.21
&
4.76
&
4.97
&
0.21
\\
\hline
\end{tabular}
\caption{ \label{tab:the}
Theoretical values in units of fermi of the
proton and neutron rms radii of
$^{133}\text{Cs}$
and
$^{127}\text{I}$
and the CsI average
obtained with
nonrelativistic Skyrme-Hartree-Fock (SHF)
and
relativistic mean field (RMF)
nuclear models.
}
\end{table*}

We fitted the COHERENT data in order to get information on the
value of the neutron rms radius $R_{n}$, which is determined by the minimization
of the $\chi^2$ in Eq.~(\ref{chi})
using the symmetrized Fermi and Helm form factors.
In both cases we obtained a minimum $\chi^2$ which is smaller than
the $\chi^2$ corresponding to full coherence by
$5.5$.
Hence,
the hypothesis of full coherence has a $p$-value of
$1.9\%$
and there is a
$2.3\sigma$
evidence of the nuclear structure suppression of the coherence.

Figure~\ref{fig:data} shows the best-fit results that we obtained
using the symmetrized Fermi and Helm form factors.
Figure~\ref{fig:rn-chi}
shows the corresponding marginal values of the $\chi^2$
as a function of $R_{n}$.
One can see from both figures that the two parameterizations of the neutron form factor
fit equally well the data
and 
give practically the same result:
\begin{equation}
R_{n}
=
5.5
{}^{+0.9}_{-1.1}
\,
\text{fm}
.
\label{rn}
\end{equation}
This is the first determination of the neutron rms radius of a nucleus
obtained with neutrino-nucleus scattering data.
Note also that it is practically model-independent,
because it coincides for the symmetrized Fermi and Helm form factors
which correspond to reasonable descriptions of the nuclear density.

\begin{figure}[!t]
\centering
\includegraphics*[width=\linewidth]{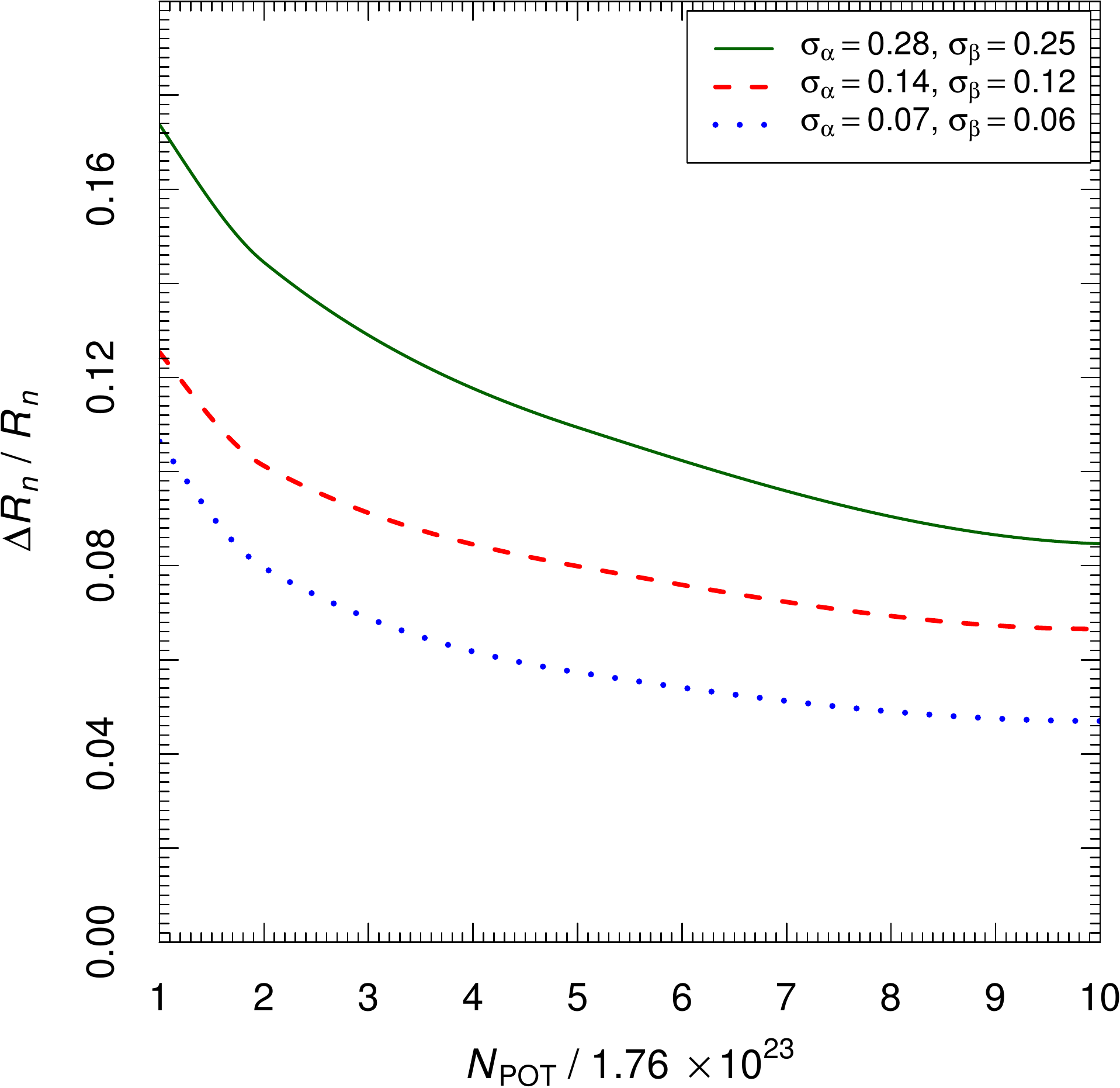}
\caption{ \label{fig:sens}
Projected relative uncertainty of the possible determination of
the neutron rms radius $R_{n}$
with the data of the COHERENT experiment
as a function of the number $N_{\text{POT}}$ of protons on target
in units of the current number
($1.76\times 10^{23}$)
for the current systematic uncertainties
(solid green curve),
half the current systematic uncertainties
(dashed red curve), and
one quarter of the current systematic uncertainties
(dotted blue curve).
}
\end{figure}

As already stated above,
the neutron rms radius was determined before only for $^{208}\text{Pb}$
from the parity-violating measurements of the PREX experiment \cite{Abrahamyan:2012gp}.
The authors of Ref.~\cite{Horowitz:2012tj} found
$
R_{n}(^{208}\text{Pb})
=
5.75 \pm 0.18
\,
\text{fm}
$.
Our best-fit value of
$R_{n}$
for
$^{127}\text{I}$ and $^{133}\text{Cs}$,
obtained assuming that the two nuclei have similar structures,
is correctly smaller than that of the heavier $^{208}\text{Pb}$ nucleus.

Table~\ref{tab:the} shows the theoretical values of the
proton and neutron rms radii of
$^{133}\text{Cs}$
and
$^{127}\text{I}$
obtained with nuclear mean field models.
All the models predict values of $R_p$ which are in approximate agreement with
the experimental ones in Eqs.~(\ref{RpCs}) and (\ref{RpI}).
Due to the large uncertainty,
the average CsI value of $R_n$ that we obtained in Eq.~(\ref{rn}) is compatible with all the model calculations.
It tends to favor values of $R_n$
that are larger than all the model calculations in Table~\ref{tab:the},
but more precise measurements are needed in order to truly test the models.

Another quantity of interest is the difference between the neutron and proton rms radii
$\Delta R_{np} = R_{n} - R_{p}$,
which is usually referred to as ``neutron skin''
\cite{Horowitz:1999fk}.
The values of $R_{p}$ for $^{127}\text{I}$ and $^{133}\text{Cs}$
determined in Ref.~\cite{Fricke:1995zz}
are around 4.78 fm, with a difference of about 0.05 fm.
Hence,
for the neutron skin, we obtain
\begin{equation}
\Delta R_{np}
\simeq
0.7
{}^{+0.9}_{-1.1}
\,
\text{fm}
.
\label{nskin}
\end{equation}
Unfortunately, the uncertainty is large and it does not allow to claim a determination
of the neutron skin.
We can only note that the best-fit value indicates the possibility of a value that is larger
than the model-predicted values in Table~\ref{tab:the},
which are between about 0.1 and 0.2 fm
(see also Ref.~\cite{Horowitz:1999fk}).

Future data of the COHERENT experiment may lead to a better determination of
the neutron rms radius $R_{n}$ and of the neutron skin $\Delta R_{np}$.
Figure~\ref{fig:sens} shows our estimation of the sensitivity to $R_{n}$ of
the COHERENT experiment as a function of the number of protons on target
with the current systematic uncertainties,
with half the current systematic uncertainties, and
with one quarter of the current systematic uncertainties.
We have included the effect of the beam-off background,
which we extracted from the statistical uncertainties of Fig.~3A of Ref.~\cite{Akimov:2017ade}.
From Fig.~\ref{fig:sens}
one can see that the current sensitivity gives a relative uncertainty
$\Delta R_{n} / R_{n} \simeq 17\%$,
which is in approximate agreement with the uncertainty of
the determination of $R_{n}$
in Eq.~(\ref{rn}).
With the current systematic uncertainties and ten times the current number of protons on target,
the data of the COHERENT experiment will allow us to determine $R_{n}$
within about 0.5 fm.
If the systematic uncertainties are reduced by half or one quarter,
$R_{n}$
can be determined within about 0.4 or 0.3 fm,
respectively.
Such a measurement would also decrease the uncertainty on the value
of the neutron skin allowing a more meaningful comparison with
the model predictions in Table~\ref{tab:the}.

Since $R_p$ is relatively well known, a measurement of $R_n$ allows to determine the neutron skin $\Delta R_{np}$. Information on this quantity is eagerly awaited because $\Delta R_{np}$ is correlated with several properties characterizing neutron-rich matter
(see Refs.~\cite{Brown:2000pd,Horowitz:2000xj,Reinhard:2010wz,Tsang:2012se,Hagen:2015yea}).
A larger neutron skin would suggest a stiffer EOS and imply a larger neutron star radius $R_{\text{NS}}$.
Since the neutron star binding energy is inversely proportional to $R_{\text{NS}}$,
a larger $R_{\text{NS}}$ implies a smaller gravitational binding energy,
which can be tested by observing the intense neutrino burst of a core collapse supernova.

The neutron skin is also correlated with several other nuclear quantities,
e.g.
with the slope of bulk symmetry energy,
with the slope of binding energy of neutron matter,
and
with the symmetry correction to the incompressibility (see Ref.~\cite{Baldo:2016jhp} for a review).

On August 17, 2017 the Advanced LIGO and Advanced Virgo gravitational-wave
detectors made their first observation of a binary neutron star inspiral \cite{TheLIGOScientific:2017qsa}. From this observation the collaboration was able to infer not only the component masses of the binary but also the tidal deformability parameter, which is related to the neutron star EOS
and to the neutron skin
\cite{Kumar:2017wqp,Fattoyev:2017jql}.

Information on the nuclear neutron density radius $R_{n}$ is also important for a precise determination of the background due to coherent elastic neutrino-nucleus scattering in dark matter detectors.
This background will crucially limit the discovery potential of future dark matter detectors~\cite{Billard:2013qya}.
Until now, the background has been evaluated using a unique Helm nuclear form factor for protons and neutrons, with the Lewin-Smith prescription~\cite{Lewin:1995rx} for the input value of the nuclear radii. 
Since Caesium and Iodine have similar atomic and mass numbers to that of Xenon,
it is possible to make an estimation of the impact of
the inclusion of different proton and neutron form factors
(with the value of $R_n$ found in this paper)
on the neutrino background for experiments like DARWIN~\cite{Aalbers:2016jon}, XENONnT~\cite{Aprile:2015uzo}, and LZ~\cite{Mount:2017qzi},
that use Xenon as a target.

In conclusion,
we have determined for the first time the neutron rms radius
of $^{133}\text{Cs}$ and $^{127}\text{I}$ (assuming that they have similar structures)
from the fit of the data on coherent elastic neutrino-nucleus scattering
of the COHERENT experiment.
Considering
the symmetrized Fermi and Helm form factors,
we obtained the practically model-independent value
$
R_{n}
=
5.5
{}^{+0.9}_{-1.1}
\,
\text{fm}
$.
We also found that the COHERENT data show a
$2.3\sigma$
evidence of the nuclear structure suppression of the full coherence.

\section*{Acknowledgment}

M. Cadeddu wishes to thank M. Lissia for useful discussions.
C. Giunti is grateful to S.M. Bilenky and M.V. Garzelli for stimulating discussions.
The work of Y.F. Li and Y.Y. Zhang was supported in part by the National Natural Science Foundation of China under Grant
No. 11305193 and by the Strategic Priority Research Program of the Chinese Academy of Sciences under Grant No. XDA10010100.
Y.F. Li is also grateful for the support by the CAS Center for Excellence in Particle Physics (CCEPP).

%

\begin{thebibliography}{43}%
\makeatletter
\providecommand \@ifxundefined [1]{%
\@ifx{#1\undefined}
}%
\providecommand \@ifnum [1]{%
\ifnum #1\expandafter \@firstoftwo
\else \expandafter \@secondoftwo
\fi
}%
\providecommand \@ifx [1]{%
\ifx #1\expandafter \@firstoftwo
\else \expandafter \@secondoftwo
\fi
}%
\providecommand \natexlab [1]{#1}%
\providecommand \enquote [1]{``#1''}%
\providecommand \bibnamefont [1]{#1}%
\providecommand \bibfnamefont [1]{#1}%
\providecommand \citenamefont [1]{#1}%
\providecommand \href@noop [0]{\@secondoftwo}%
\providecommand \href [0]{\begingroup \@sanitize@url \@href}%
\providecommand \@href[1]{\@@startlink{#1}\@@href}%
\providecommand \@@href[1]{\endgroup#1\@@endlink}%
\providecommand \@sanitize@url [0]{\catcode `\\12\catcode `\$12\catcode
`\&12\catcode `\#12\catcode `\^12\catcode `\_12\catcode `\%12\relax}%
\providecommand \@@startlink[1]{}%
\providecommand \@@endlink[0]{}%
\providecommand \url [0]{\begingroup\@sanitize@url \@url }%
\providecommand \@url [1]{\endgroup\@href {#1}{\urlprefix }}%
\providecommand \urlprefix [0]{URL }%
\providecommand \Eprint [0]{\href }%
\providecommand \doibase [0]{http://dx.doi.org/}%
\providecommand \selectlanguage [0]{\@gobble}%
\providecommand \bibinfo [0]{\@secondoftwo}%
\providecommand \bibfield [0]{\@secondoftwo}%
\providecommand \translation [1]{[#1]}%
\providecommand \BibitemOpen [0]{}%
\providecommand \bibitemStop [0]{}%
\providecommand \bibitemNoStop [0]{.\EOS\space}%
\providecommand \EOS [0]{\spacefactor3000\relax}%
\providecommand \BibitemShut [1]{\csname bibitem#1\endcsname}%
\let\auto@bib@innerbib\@empty
\bibitem [{\citenamefont {Akimov}\ \emph {et~al.}(2017)\citenamefont {Akimov}
\emph {et~al.}}]{Akimov:2017ade}%
\BibitemOpen
\bibfield {author} {\bibinfo {author} {\bibfnamefont {D.}~\bibnamefont
{Akimov}} \emph {et~al.} (\bibinfo {collaboration} {COHERENT}),\ }\href
{\doibase 10.1126/science.aao0990} {\bibfield {journal} {\bibinfo {journal}
{Science}\ }\textbf {\bibinfo {volume} {357}},\ \bibinfo {pages} {1123}
(\bibinfo {year} {2017})},\ \Eprint {http://arxiv.org/abs/arXiv:1708.01294}
{arXiv:1708.01294 [nucl-ex]} \BibitemShut {NoStop}%
\bibitem [{\citenamefont {Freedman}(1974)}]{Freedman:1973yd}%
\BibitemOpen
\bibfield {author} {\bibinfo {author} {\bibfnamefont {D.~Z.}\ \bibnamefont
{Freedman}},\ }\href {\doibase 10.1103/PhysRevD.9.1389} {\bibfield {journal}
{\bibinfo {journal} {Phys. Rev.}\ }\textbf {\bibinfo {volume} {D9}},\
\bibinfo {pages} {1389} (\bibinfo {year} {1974})}\BibitemShut {NoStop}%
\bibitem [{\citenamefont {Freedman}\ \emph {et~al.}(1977)\citenamefont
{Freedman}, \citenamefont {Schramm},\ and\ \citenamefont
{Tubbs}}]{Freedman:1977xn}%
\BibitemOpen
\bibfield {author} {\bibinfo {author} {\bibfnamefont {D.~Z.}\ \bibnamefont
{Freedman}}, \bibinfo {author} {\bibfnamefont {D.~N.}\ \bibnamefont
{Schramm}}, \ and\ \bibinfo {author} {\bibfnamefont {D.~L.}\ \bibnamefont
{Tubbs}},\ }\href {\doibase 10.1146/annurev.ns.27.120177.001123} {\bibfield
{journal} {\bibinfo {journal} {Ann. Rev. Nucl. Part. Sci.}\ }\textbf
{\bibinfo {volume} {27}},\ \bibinfo {pages} {167} (\bibinfo {year}
{1977})}\BibitemShut {NoStop}%
\bibitem [{\citenamefont {Drukier}\ and\ \citenamefont
{Stodolsky}(1984)}]{Drukier:1983gj}%
\BibitemOpen
\bibfield {author} {\bibinfo {author} {\bibfnamefont {A.}~\bibnamefont
{Drukier}}\ and\ \bibinfo {author} {\bibfnamefont {L.}~\bibnamefont
{Stodolsky}},\ }\href {\doibase 10.1103/PhysRevD.30.2295} {\bibfield
{journal} {\bibinfo {journal} {Phys. Rev.}\ }\textbf {\bibinfo {volume}
{D30}},\ \bibinfo {pages} {2295} (\bibinfo {year} {1984})}\BibitemShut
{NoStop}%
\bibitem [{\citenamefont {Barranco}\ \emph {et~al.}(2005)\citenamefont
{Barranco}, \citenamefont {Miranda},\ and\ \citenamefont
{Rashba}}]{Barranco:2005yy}%
\BibitemOpen
\bibfield {author} {\bibinfo {author} {\bibfnamefont {J.}~\bibnamefont
{Barranco}}, \bibinfo {author} {\bibfnamefont {O.~G.}\ \bibnamefont
{Miranda}}, \ and\ \bibinfo {author} {\bibfnamefont {T.~I.}\ \bibnamefont
{Rashba}},\ }\href@noop {} {\bibfield {journal} {\bibinfo {journal} {JHEP}\
}\textbf {\bibinfo {volume} {0512}},\ \bibinfo {pages} {021} (\bibinfo {year}
{2005})},\ \Eprint {http://arxiv.org/abs/hep-ph/0508299} {hep-ph/0508299}
\BibitemShut {NoStop}%
\bibitem [{\citenamefont {Patton}\ \emph {et~al.}(2012)\citenamefont {Patton},
\citenamefont {Engel}, \citenamefont {McLaughlin},\ and\ \citenamefont
{Schunck}}]{Patton:2012jr}%
\BibitemOpen
\bibfield {author} {\bibinfo {author} {\bibfnamefont {K.}~\bibnamefont
{Patton}}, \bibinfo {author} {\bibfnamefont {J.}~\bibnamefont {Engel}},
\bibinfo {author} {\bibfnamefont {G.~C.}\ \bibnamefont {McLaughlin}}, \ and\
\bibinfo {author} {\bibfnamefont {N.}~\bibnamefont {Schunck}},\ }\href
{\doibase 10.1103/PhysRevC.86.024612} {\bibfield {journal} {\bibinfo
{journal} {Phys. Rev.}\ }\textbf {\bibinfo {volume} {C86}},\ \bibinfo {pages}
{024612} (\bibinfo {year} {2012})},\ \Eprint
{http://arxiv.org/abs/arXiv:1207.0693} {arXiv:1207.0693 [nucl-th]}
\BibitemShut {NoStop}%
\bibitem [{\citenamefont {Papoulias}\ and\ \citenamefont
{Kosmas}(2015)}]{Papoulias:2015vxa}%
\BibitemOpen
\bibfield {author} {\bibinfo {author} {\bibfnamefont {D.~K.}\ \bibnamefont
{Papoulias}}\ and\ \bibinfo {author} {\bibfnamefont {T.~S.}\ \bibnamefont
{Kosmas}},\ }\href {\doibase 10.1155/2015/763648} {\bibfield {journal}
{\bibinfo {journal} {Adv. High Energy Phys.}\ }\textbf {\bibinfo {volume}
{2015}},\ \bibinfo {pages} {763648} (\bibinfo {year} {2015})},\ \Eprint
{http://arxiv.org/abs/arXiv:1502.02928} {arXiv:1502.02928 [nucl-th]}
\BibitemShut {NoStop}%
\bibitem [{\citenamefont {Patrignani}\ \emph {et~al.}(2016)\citenamefont
{Patrignani} \emph {et~al.}}]{PDG-2016}%
\BibitemOpen
\bibfield {author} {\bibinfo {author} {\bibfnamefont {C.}~\bibnamefont
{Patrignani}} \emph {et~al.} (\bibinfo {collaboration} {Particle Data
Group}),\ }\href {\doibase 10.1088/1674-1137/40/10/100001} {\bibfield
{journal} {\bibinfo {journal} {Chin. Phys.}\ }\textbf {\bibinfo {volume}
{C40}},\ \bibinfo {pages} {100001} (\bibinfo {year} {2016})}\BibitemShut
{NoStop}%
\bibitem [{\citenamefont {Fricke}\ \emph {et~al.}(1995)\citenamefont {Fricke},
\citenamefont {Bernhardt}, \citenamefont {Heilig}, \citenamefont {Schaller},
\citenamefont {Schellenberg}, \citenamefont {Shera},\ and\ \citenamefont
{de~Jager}}]{Fricke:1995zz}%
\BibitemOpen
\bibfield {author} {\bibinfo {author} {\bibfnamefont {G.}~\bibnamefont
{Fricke}}, \bibinfo {author} {\bibfnamefont {C.}~\bibnamefont {Bernhardt}},
\bibinfo {author} {\bibfnamefont {K.}~\bibnamefont {Heilig}}, \bibinfo
{author} {\bibfnamefont {L.~A.}\ \bibnamefont {Schaller}}, \bibinfo {author}
{\bibfnamefont {L.}~\bibnamefont {Schellenberg}}, \bibinfo {author}
{\bibfnamefont {E.~B.}\ \bibnamefont {Shera}}, \ and\ \bibinfo {author}
{\bibfnamefont {C.~W.}\ \bibnamefont {de~Jager}},\ }\href {\doibase
10.1006/adnd.1995.1007} {\bibfield {journal} {\bibinfo {journal} {Atom.
Data Nucl. Data Tabl.}\ }\textbf {\bibinfo {volume} {60}},\ \bibinfo {pages}
{177} (\bibinfo {year} {1995})}\BibitemShut {NoStop}%
\bibitem [{\citenamefont {Angeli}\ and\ \citenamefont
{Marinova}(2013)}]{Angeli:2013epw}%
\BibitemOpen
\bibfield {author} {\bibinfo {author} {\bibfnamefont {I.}~\bibnamefont
{Angeli}}\ and\ \bibinfo {author} {\bibfnamefont {K.~P.}\ \bibnamefont
{Marinova}},\ }\href {\doibase 10.1016/j.adt.2011.12.006} {\bibfield
{journal} {\bibinfo {journal} {Atom. Data Nucl. Data Tabl.}\ }\textbf
{\bibinfo {volume} {99}},\ \bibinfo {pages} {69} (\bibinfo {year}
{2013})}\BibitemShut {NoStop}%
\bibitem [{\citenamefont {Garcia-Recio}\ \emph {et~al.}(1992)\citenamefont
{Garcia-Recio}, \citenamefont {Nieves},\ and\ \citenamefont
{Oset}}]{GarciaRecio:1991wk}%
\BibitemOpen
\bibfield {author} {\bibinfo {author} {\bibfnamefont {C.}~\bibnamefont
{Garcia-Recio}}, \bibinfo {author} {\bibfnamefont {J.}~\bibnamefont
{Nieves}}, \ and\ \bibinfo {author} {\bibfnamefont {E.}~\bibnamefont
{Oset}},\ }\href {\doibase 10.1016/0375-9474(92)90034-H} {\bibfield
{journal} {\bibinfo {journal} {Nucl. Phys.}\ }\textbf {\bibinfo {volume}
{A547}},\ \bibinfo {pages} {473} (\bibinfo {year} {1992})}\BibitemShut
{NoStop}%
\bibitem [{\citenamefont {Starodubsky}\ and\ \citenamefont
{Hintz}(1994)}]{Starodubsky:1994xt}%
\BibitemOpen
\bibfield {author} {\bibinfo {author} {\bibfnamefont {V.~E.}\ \bibnamefont
{Starodubsky}}\ and\ \bibinfo {author} {\bibfnamefont {N.~M.}\ \bibnamefont
{Hintz}},\ }\href {\doibase 10.1103/PhysRevC.49.2118} {\bibfield {journal}
{\bibinfo {journal} {Phys. Rev.}\ }\textbf {\bibinfo {volume} {C49}},\
\bibinfo {pages} {2118} (\bibinfo {year} {1994})}\BibitemShut {NoStop}%
\bibitem [{\citenamefont {Trzcinska}\ \emph {et~al.}(2001)\citenamefont
{Trzcinska}, \citenamefont {Jastrzebski}, \citenamefont {Lubinski},
\citenamefont {Hartmann}, \citenamefont {Schmidt}, \citenamefont {von
Egidy},\ and\ \citenamefont {Klos}}]{Trzcinska:2001sy}%
\BibitemOpen
\bibfield {author} {\bibinfo {author} {\bibfnamefont {A.}~\bibnamefont
{Trzcinska}}, \bibinfo {author} {\bibfnamefont {J.}~\bibnamefont
{Jastrzebski}}, \bibinfo {author} {\bibfnamefont {P.}~\bibnamefont
{Lubinski}}, \bibinfo {author} {\bibfnamefont {F.~J.}\ \bibnamefont
{Hartmann}}, \bibinfo {author} {\bibfnamefont {R.}~\bibnamefont {Schmidt}},
\bibinfo {author} {\bibfnamefont {T.}~\bibnamefont {von Egidy}}, \ and\
\bibinfo {author} {\bibfnamefont {B.}~\bibnamefont {Klos}},\ }\href {\doibase
10.1103/PhysRevLett.87.082501} {\bibfield {journal} {\bibinfo {journal}
{Phys. Rev. Lett.}\ }\textbf {\bibinfo {volume} {87}},\ \bibinfo {pages}
{082501} (\bibinfo {year} {2001})}\BibitemShut {NoStop}%
\bibitem [{\citenamefont {Clark}\ \emph {et~al.}(2003)\citenamefont {Clark},
\citenamefont {Kerr},\ and\ \citenamefont {Hama}}]{Clark:2002se}%
\BibitemOpen
\bibfield {author} {\bibinfo {author} {\bibfnamefont {B.~C.}\ \bibnamefont
{Clark}}, \bibinfo {author} {\bibfnamefont {L.~J.}\ \bibnamefont {Kerr}}, \
and\ \bibinfo {author} {\bibfnamefont {S.}~\bibnamefont {Hama}},\ }\href
{\doibase 10.1103/PhysRevC.67.054605} {\bibfield {journal} {\bibinfo
{journal} {Phys. Rev.}\ }\textbf {\bibinfo {volume} {C67}},\ \bibinfo {pages}
{054605} (\bibinfo {year} {2003})},\ \Eprint
{http://arxiv.org/abs/nucl-th/0209052} {nucl-th/0209052 [nucl-th]}
\BibitemShut {NoStop}%
\bibitem [{\citenamefont {Abrahamyan}\ \emph {et~al.}(2012)\citenamefont
{Abrahamyan} \emph {et~al.}}]{Abrahamyan:2012gp}%
\BibitemOpen
\bibfield {author} {\bibinfo {author} {\bibfnamefont {S.}~\bibnamefont
{Abrahamyan}} \emph {et~al.} (\bibinfo {collaboration} {PREX}),\ }\href
{\doibase 10.1103/PhysRevLett.108.112502} {\bibfield {journal} {\bibinfo
{journal} {Phys. Rev. Lett.}\ }\textbf {\bibinfo {volume} {108}},\ \bibinfo
{pages} {112502} (\bibinfo {year} {2012})},\ \Eprint
{http://arxiv.org/abs/arXiv:1201.2568} {arXiv:1201.2568 [nucl-ex]}
\BibitemShut {NoStop}%
\bibitem [{\citenamefont {Brown}(2000)}]{Brown:2000pd}%
\BibitemOpen
\bibfield {author} {\bibinfo {author} {\bibfnamefont {B.~A.}\ \bibnamefont
{Brown}},\ }\href {\doibase 10.1103/PhysRevLett.85.5296} {\bibfield
{journal} {\bibinfo {journal} {Phys. Rev. Lett.}\ }\textbf {\bibinfo
{volume} {85}},\ \bibinfo {pages} {5296} (\bibinfo {year}
{2000})}\BibitemShut {NoStop}%
\bibitem [{\citenamefont {Horowitz}\ and\ \citenamefont
{Piekarewicz}(2001)}]{Horowitz:2000xj}%
\BibitemOpen
\bibfield {author} {\bibinfo {author} {\bibfnamefont {C.~J.}\ \bibnamefont
{Horowitz}}\ and\ \bibinfo {author} {\bibfnamefont {J.}~\bibnamefont
{Piekarewicz}},\ }\href {\doibase 10.1103/PhysRevLett.86.5647} {\bibfield
{journal} {\bibinfo {journal} {Phys. Rev. Lett.}\ }\textbf {\bibinfo
{volume} {86}},\ \bibinfo {pages} {5647} (\bibinfo {year} {2001})},\ \Eprint
{http://arxiv.org/abs/astro-ph/0010227} {astro-ph/0010227 [astro-ph]}
\BibitemShut {NoStop}%
\bibitem [{\citenamefont {Reinhard}\ and\ \citenamefont
{Nazarewicz}(2010)}]{Reinhard:2010wz}%
\BibitemOpen
\bibfield {author} {\bibinfo {author} {\bibfnamefont {P.~G.}\ \bibnamefont
{Reinhard}}\ and\ \bibinfo {author} {\bibfnamefont {W.}~\bibnamefont
{Nazarewicz}},\ }\href {\doibase 10.1103/PhysRevC.81.051303} {\bibfield
{journal} {\bibinfo {journal} {Phys. Rev.}\ }\textbf {\bibinfo {volume}
{C81}},\ \bibinfo {pages} {051303} (\bibinfo {year} {2010})},\ \Eprint
{http://arxiv.org/abs/arXiv:1002.4140} {arXiv:1002.4140 [nucl-th]}
\BibitemShut {NoStop}%
\bibitem [{\citenamefont {Tsang}\ \emph {et~al.}(2012)\citenamefont {Tsang}
\emph {et~al.}}]{Tsang:2012se}%
\BibitemOpen
\bibfield {author} {\bibinfo {author} {\bibfnamefont {M.~B.}\ \bibnamefont
{Tsang}} \emph {et~al.},\ }\href {\doibase 10.1103/PhysRevC.86.015803}
{\bibfield {journal} {\bibinfo {journal} {Phys. Rev.}\ }\textbf {\bibinfo
{volume} {C86}},\ \bibinfo {pages} {015803} (\bibinfo {year} {2012})},\
\Eprint {http://arxiv.org/abs/arXiv:1204.0466} {arXiv:1204.0466 [nucl-ex]}
\BibitemShut {NoStop}%
\bibitem [{\citenamefont {Hagen}\ \emph {et~al.}(2015)\citenamefont {Hagen}
\emph {et~al.}}]{Hagen:2015yea}%
\BibitemOpen
\bibfield {author} {\bibinfo {author} {\bibfnamefont {G.}~\bibnamefont
{Hagen}} \emph {et~al.},\ }\href {\doibase 10.1038/nphys3529} {\bibfield
{journal} {\bibinfo {journal} {Nature Phys.}\ }\textbf {\bibinfo {volume}
{12}},\ \bibinfo {pages} {186} (\bibinfo {year} {2015})},\ \Eprint
{http://arxiv.org/abs/arXiv:1509.07169} {arXiv:1509.07169 [nucl-th]}
\BibitemShut {NoStop}%
\bibitem [{\citenamefont {Piekarewicz}\ \emph {et~al.}(2016)\citenamefont
{Piekarewicz}, \citenamefont {Linero}, \citenamefont {Giuliani},\ and\
\citenamefont {Chicken}}]{Piekarewicz:2016vbn}%
\BibitemOpen
\bibfield {author} {\bibinfo {author} {\bibfnamefont {J.}~\bibnamefont
{Piekarewicz}}, \bibinfo {author} {\bibfnamefont {A.~R.}\ \bibnamefont
{Linero}}, \bibinfo {author} {\bibfnamefont {P.}~\bibnamefont {Giuliani}}, \
and\ \bibinfo {author} {\bibfnamefont {E.}~\bibnamefont {Chicken}},\ }\href
{\doibase 10.1103/PhysRevC.94.034316} {\bibfield {journal} {\bibinfo
{journal} {Phys. Rev.}\ }\textbf {\bibinfo {volume} {C94}},\ \bibinfo {pages}
{034316} (\bibinfo {year} {2016})},\ \Eprint
{http://arxiv.org/abs/arXiv:1604.07799} {arXiv:1604.07799 [nucl-th]}
\BibitemShut {NoStop}%
\bibitem [{\citenamefont {Helm}(1956)}]{Helm:1956zz}%
\BibitemOpen
\bibfield {author} {\bibinfo {author} {\bibfnamefont {R.~H.}\ \bibnamefont
{Helm}},\ }\href {\doibase 10.1103/PhysRev.104.1466} {\bibfield {journal}
{\bibinfo {journal} {Phys. Rev.}\ }\textbf {\bibinfo {volume} {104}},\
\bibinfo {pages} {1466} (\bibinfo {year} {1956})}\BibitemShut {NoStop}%
\bibitem [{\citenamefont {Friedrich}\ and\ \citenamefont
{Voegler}(1982)}]{Friedrich:1982esq}%
\BibitemOpen
\bibfield {author} {\bibinfo {author} {\bibfnamefont {J.}~\bibnamefont
{Friedrich}}\ and\ \bibinfo {author} {\bibfnamefont {N.}~\bibnamefont
{Voegler}},\ }\href {\doibase 10.1016/0375-9474(82)90147-6} {\bibfield
{journal} {\bibinfo {journal} {Nucl. Phys.}\ }\textbf {\bibinfo {volume}
{A373}},\ \bibinfo {pages} {192} (\bibinfo {year} {1982})}\BibitemShut
{NoStop}%
\bibitem [{\citenamefont {Klein}\ and\ \citenamefont
{Nystrand}(1999)}]{Klein:1999qj}%
\BibitemOpen
\bibfield {author} {\bibinfo {author} {\bibfnamefont {S.}~\bibnamefont
{Klein}}\ and\ \bibinfo {author} {\bibfnamefont {J.}~\bibnamefont
{Nystrand}},\ }\href {\doibase 10.1103/PhysRevC.60.014903} {\bibfield
{journal} {\bibinfo {journal} {Phys. Rev.}\ }\textbf {\bibinfo {volume}
{C60}},\ \bibinfo {pages} {014903} (\bibinfo {year} {1999})},\ \Eprint
{http://arxiv.org/abs/hep-ph/9902259} {hep-ph/9902259 [hep-ph]} \BibitemShut
{NoStop}%
\bibitem [{\citenamefont {Bartel}\ \emph {et~al.}(1982)\citenamefont {Bartel},
\citenamefont {Quentin}, \citenamefont {Brack}, \citenamefont {Guet},\ and\
\citenamefont {Hakansson}}]{Bartel:1982ed}%
\BibitemOpen
\bibfield {author} {\bibinfo {author} {\bibfnamefont {J.}~\bibnamefont
{Bartel}}, \bibinfo {author} {\bibfnamefont {P.}~\bibnamefont {Quentin}},
\bibinfo {author} {\bibfnamefont {M.}~\bibnamefont {Brack}}, \bibinfo
{author} {\bibfnamefont {C.}~\bibnamefont {Guet}}, \ and\ \bibinfo {author}
{\bibfnamefont {H.~B.}\ \bibnamefont {Hakansson}},\ }\href {\doibase
10.1016/0375-9474(82)90403-1} {\bibfield {journal} {\bibinfo {journal}
{Nucl. Phys.}\ }\textbf {\bibinfo {volume} {A386}},\ \bibinfo {pages} {79}
(\bibinfo {year} {1982})}\BibitemShut {NoStop}%
\bibitem [{\citenamefont {Dobaczewski}\ \emph {et~al.}(1984)\citenamefont
{Dobaczewski}, \citenamefont {Flocard},\ and\ \citenamefont
{Treiner}}]{Dobaczewski:1983zc}%
\BibitemOpen
\bibfield {author} {\bibinfo {author} {\bibfnamefont {J.}~\bibnamefont
{Dobaczewski}}, \bibinfo {author} {\bibfnamefont {H.}~\bibnamefont
{Flocard}}, \ and\ \bibinfo {author} {\bibfnamefont {J.}~\bibnamefont
{Treiner}},\ }\href {\doibase 10.1016/0375-9474(84)90433-0} {\bibfield
{journal} {\bibinfo {journal} {Nucl. Phys.}\ }\textbf {\bibinfo {volume}
{A422}},\ \bibinfo {pages} {103} (\bibinfo {year} {1984})}\BibitemShut
{NoStop}%
\bibitem [{\citenamefont {Reinhard}\ and\ \citenamefont
{Flocard}(1995)}]{Reinhard:1995zz}%
\BibitemOpen
\bibfield {author} {\bibinfo {author} {\bibfnamefont {P.~G.}\ \bibnamefont
{Reinhard}}\ and\ \bibinfo {author} {\bibfnamefont {H.}~\bibnamefont
{Flocard}},\ }\href {\doibase 10.1016/0375-9474(94)00770-N} {\bibfield
{journal} {\bibinfo {journal} {Nucl. Phys.}\ }\textbf {\bibinfo {volume}
{A584}},\ \bibinfo {pages} {467} (\bibinfo {year} {1995})}\BibitemShut
{NoStop}%
\bibitem [{\citenamefont {Chabanat}\ \emph {et~al.}(1998)\citenamefont
{Chabanat}, \citenamefont {Bonche}, \citenamefont {Haensel}, \citenamefont
{Meyer},\ and\ \citenamefont {Schaeffer}}]{Chabanat:1997un}%
\BibitemOpen
\bibfield {author} {\bibinfo {author} {\bibfnamefont {E.}~\bibnamefont
{Chabanat}}, \bibinfo {author} {\bibfnamefont {P.}~\bibnamefont {Bonche}},
\bibinfo {author} {\bibfnamefont {P.}~\bibnamefont {Haensel}}, \bibinfo
{author} {\bibfnamefont {J.}~\bibnamefont {Meyer}}, \ and\ \bibinfo {author}
{\bibfnamefont {R.}~\bibnamefont {Schaeffer}},\ }\href {\doibase
10.1016/S0375-9474(98)00570-3, 10.1016/S0375-9474(98)00180-8} {\bibfield
{journal} {\bibinfo {journal} {Nucl. Phys.}\ }\textbf {\bibinfo {volume}
{A635}},\ \bibinfo {pages} {231} (\bibinfo {year} {1998})}\BibitemShut
{NoStop}%
\bibitem [{\citenamefont {Kortelainen}\ \emph {et~al.}(2012)\citenamefont
{Kortelainen}, \citenamefont {McDonnell}, \citenamefont {Nazarewicz},
\citenamefont {Reinhard}, \citenamefont {Sarich}, \citenamefont {Schunck},
\citenamefont {Stoitsov},\ and\ \citenamefont {Wild}}]{Kortelainen:2011ft}%
\BibitemOpen
\bibfield {author} {\bibinfo {author} {\bibfnamefont {M.}~\bibnamefont
{Kortelainen}}, \bibinfo {author} {\bibfnamefont {J.}~\bibnamefont
{McDonnell}}, \bibinfo {author} {\bibfnamefont {W.}~\bibnamefont
{Nazarewicz}}, \bibinfo {author} {\bibfnamefont {P.~G.}\ \bibnamefont
{Reinhard}}, \bibinfo {author} {\bibfnamefont {J.}~\bibnamefont {Sarich}},
\bibinfo {author} {\bibfnamefont {N.}~\bibnamefont {Schunck}}, \bibinfo
{author} {\bibfnamefont {M.~V.}\ \bibnamefont {Stoitsov}}, \ and\ \bibinfo
{author} {\bibfnamefont {S.~M.}\ \bibnamefont {Wild}},\ }\href {\doibase
10.1103/PhysRevC.85.024304} {\bibfield {journal} {\bibinfo {journal} {Phys.
Rev.}\ }\textbf {\bibinfo {volume} {C85}},\ \bibinfo {pages} {024304}
(\bibinfo {year} {2012})},\ \Eprint {http://arxiv.org/abs/arXiv:1111.4344}
{arXiv:1111.4344 [nucl-th]} \BibitemShut {NoStop}%
\bibitem [{\citenamefont {Sharma}\ \emph {et~al.}(1993)\citenamefont {Sharma},
\citenamefont {Nagarajan},\ and\ \citenamefont {Ring}}]{Sharma:1993it}%
\BibitemOpen
\bibfield {author} {\bibinfo {author} {\bibfnamefont {M.~M.}\ \bibnamefont
{Sharma}}, \bibinfo {author} {\bibfnamefont {M.~A.}\ \bibnamefont
{Nagarajan}}, \ and\ \bibinfo {author} {\bibfnamefont {P.}~\bibnamefont
{Ring}},\ }\href {\doibase 10.1016/0370-2693(93)90970-S} {\bibfield
{journal} {\bibinfo {journal} {Phys. Lett.}\ }\textbf {\bibinfo {volume}
{B312}},\ \bibinfo {pages} {377} (\bibinfo {year} {1993})}\BibitemShut
{NoStop}%
\bibitem [{\citenamefont {Lalazissis}\ \emph {et~al.}(1997)\citenamefont
{Lalazissis}, \citenamefont {Konig},\ and\ \citenamefont
{Ring}}]{Lalazissis:1996rd}%
\BibitemOpen
\bibfield {author} {\bibinfo {author} {\bibfnamefont {G.~A.}\ \bibnamefont
{Lalazissis}}, \bibinfo {author} {\bibfnamefont {J.}~\bibnamefont {Konig}}, \
and\ \bibinfo {author} {\bibfnamefont {P.}~\bibnamefont {Ring}},\ }\href
{\doibase 10.1103/PhysRevC.55.540} {\bibfield {journal} {\bibinfo {journal}
{Phys. Rev.}\ }\textbf {\bibinfo {volume} {C55}},\ \bibinfo {pages} {540}
(\bibinfo {year} {1997})},\ \Eprint {http://arxiv.org/abs/nucl-th/9607039}
{nucl-th/9607039 [nucl-th]} \BibitemShut {NoStop}%
\bibitem [{\citenamefont {Bender}\ \emph {et~al.}(1999)\citenamefont {Bender},
\citenamefont {Rutz}, \citenamefont {Reinhard}, \citenamefont {Maruhn},\ and\
\citenamefont {Greiner}}]{Bender:1999yt}%
\BibitemOpen
\bibfield {author} {\bibinfo {author} {\bibfnamefont {M.}~\bibnamefont
{Bender}}, \bibinfo {author} {\bibfnamefont {K.}~\bibnamefont {Rutz}},
\bibinfo {author} {\bibfnamefont {P.~G.}\ \bibnamefont {Reinhard}}, \bibinfo
{author} {\bibfnamefont {J.~A.}\ \bibnamefont {Maruhn}}, \ and\ \bibinfo
{author} {\bibfnamefont {W.}~\bibnamefont {Greiner}},\ }\href {\doibase
10.1103/PhysRevC.60.034304} {\bibfield {journal} {\bibinfo {journal} {Phys.
Rev.}\ }\textbf {\bibinfo {volume} {C60}},\ \bibinfo {pages} {034304}
(\bibinfo {year} {1999})},\ \Eprint {http://arxiv.org/abs/nucl-th/9906030}
{nucl-th/9906030 [nucl-th]} \BibitemShut {NoStop}%
\bibitem [{\citenamefont {Horowitz}\ \emph {et~al.}(2012)\citenamefont
{Horowitz} \emph {et~al.}}]{Horowitz:2012tj}%
\BibitemOpen
\bibfield {author} {\bibinfo {author} {\bibfnamefont {C.~J.}\ \bibnamefont
{Horowitz}} \emph {et~al.},\ }\href {\doibase 10.1103/PhysRevC.85.032501}
{\bibfield {journal} {\bibinfo {journal} {Phys. Rev.}\ }\textbf {\bibinfo
{volume} {C85}},\ \bibinfo {pages} {032501} (\bibinfo {year} {2012})},\
\Eprint {http://arxiv.org/abs/arXiv:1202.1468} {arXiv:1202.1468 [nucl-ex]}
\BibitemShut {NoStop}%
\bibitem [{\citenamefont {Horowitz}\ \emph {et~al.}(2001)\citenamefont
{Horowitz}, \citenamefont {Pollock}, \citenamefont {Souder},\ and\
\citenamefont {Michaels}}]{Horowitz:1999fk}%
\BibitemOpen
\bibfield {author} {\bibinfo {author} {\bibfnamefont {C.~J.}\ \bibnamefont
{Horowitz}}, \bibinfo {author} {\bibfnamefont {S.~J.}\ \bibnamefont
{Pollock}}, \bibinfo {author} {\bibfnamefont {P.~A.}\ \bibnamefont {Souder}},
\ and\ \bibinfo {author} {\bibfnamefont {R.}~\bibnamefont {Michaels}},\
}\href {\doibase 10.1103/PhysRevC.63.025501} {\bibfield {journal} {\bibinfo
{journal} {Phys. Rev.}\ }\textbf {\bibinfo {volume} {C63}},\ \bibinfo {pages}
{025501} (\bibinfo {year} {2001})},\ \Eprint
{http://arxiv.org/abs/nucl-th/9912038} {nucl-th/9912038 [nucl-th]}
\BibitemShut {NoStop}%
\bibitem [{\citenamefont {Baldo}\ and\ \citenamefont
{Burgio}(2016)}]{Baldo:2016jhp}%
\BibitemOpen
\bibfield {author} {\bibinfo {author} {\bibfnamefont {M.}~\bibnamefont
{Baldo}}\ and\ \bibinfo {author} {\bibfnamefont {G.~F.}\ \bibnamefont
{Burgio}},\ }\href@noop {} {\bibfield {journal} {\bibinfo {journal}
{Prog.Part.Nucl. Phys.}\ }\textbf {\bibinfo {volume} {91}},\ \bibinfo {pages}
{203} (\bibinfo {year} {2016})},\ \Eprint
{http://arxiv.org/abs/arXiv:1606.08838} {arXiv:1606.08838 [nucl-th]}
\BibitemShut {NoStop}%
\bibitem [{\citenamefont {Abbott}\ \emph {et~al.}(2017)\citenamefont {Abbott}
\emph {et~al.}}]{TheLIGOScientific:2017qsa}%
\BibitemOpen
\bibfield {author} {\bibinfo {author} {\bibfnamefont {B.~P.}\ \bibnamefont
{Abbott}} \emph {et~al.} (\bibinfo {collaboration} {Virgo, LIGO
Scientific}),\ }\href {\doibase 10.1103/PhysRevLett.119.161101} {\bibfield
{journal} {\bibinfo {journal} {Phys. Rev. Lett.}\ }\textbf {\bibinfo
{volume} {119}},\ \bibinfo {pages} {161101} (\bibinfo {year} {2017})},\
\Eprint {http://arxiv.org/abs/arXiv:1710.05832} {arXiv:1710.05832 [gr-qc]}
\BibitemShut {NoStop}%
\bibitem [{\citenamefont {Kumar}\ \emph {et~al.}()\citenamefont {Kumar},
\citenamefont {Agrawal},\ and\ \citenamefont {Patra}}]{Kumar:2017wqp}%
\BibitemOpen
\bibfield {author} {\bibinfo {author} {\bibfnamefont {B.}~\bibnamefont
{Kumar}}, \bibinfo {author} {\bibfnamefont {B.~K.}\ \bibnamefont {Agrawal}},
\ and\ \bibinfo {author} {\bibfnamefont {S.~K.}\ \bibnamefont {Patra}},\
}\href@noop {} {\ }\Eprint {http://arxiv.org/abs/arXiv:1711.04940}
{arXiv:1711.04940 [nucl-th]} \BibitemShut {NoStop}%
\bibitem [{\citenamefont {Fattoyev}\ \emph {et~al.}()\citenamefont {Fattoyev},
\citenamefont {Piekarewicz},\ and\ \citenamefont
{Horowitz}}]{Fattoyev:2017jql}%
\BibitemOpen
\bibfield {author} {\bibinfo {author} {\bibfnamefont {F.~J.}\ \bibnamefont
{Fattoyev}}, \bibinfo {author} {\bibfnamefont {J.}~\bibnamefont
{Piekarewicz}}, \ and\ \bibinfo {author} {\bibfnamefont {C.~J.}\ \bibnamefont
{Horowitz}},\ }\href@noop {} {\ }\Eprint
{http://arxiv.org/abs/arXiv:1711.06615} {arXiv:1711.06615 [nucl-th]}
\BibitemShut {NoStop}%
\bibitem [{\citenamefont {Billard}\ \emph {et~al.}(2014)\citenamefont
{Billard}, \citenamefont {Strigari},\ and\ \citenamefont
{Figueroa-Feliciano}}]{Billard:2013qya}%
\BibitemOpen
\bibfield {author} {\bibinfo {author} {\bibfnamefont {J.}~\bibnamefont
{Billard}}, \bibinfo {author} {\bibfnamefont {L.}~\bibnamefont {Strigari}}, \
and\ \bibinfo {author} {\bibfnamefont {E.}~\bibnamefont
{Figueroa-Feliciano}},\ }\href@noop {} {\bibfield {journal} {\bibinfo
{journal} {Phys. Rev.}\ }\textbf {\bibinfo {volume} {D89}},\ \bibinfo {pages}
{023524} (\bibinfo {year} {2014})},\ \Eprint
{http://arxiv.org/abs/arXiv:1307.5458} {arXiv:1307.5458 [hep-ph]}
\BibitemShut {NoStop}%
\bibitem [{\citenamefont {Lewin}\ and\ \citenamefont
{Smith}(1996)}]{Lewin:1995rx}%
\BibitemOpen
\bibfield {author} {\bibinfo {author} {\bibfnamefont {J.~D.}\ \bibnamefont
{Lewin}}\ and\ \bibinfo {author} {\bibfnamefont {P.~F.}\ \bibnamefont
{Smith}},\ }\href {\doibase 10.1016/S0927-6505(96)00047-3} {\bibfield
{journal} {\bibinfo {journal} {Astropart. Phys.}\ }\textbf {\bibinfo
{volume} {6}},\ \bibinfo {pages} {87} (\bibinfo {year} {1996})}\BibitemShut
{NoStop}%
\bibitem [{\citenamefont {Aalbers}\ \emph {et~al.}(2016)\citenamefont {Aalbers}
\emph {et~al.}}]{Aalbers:2016jon}%
\BibitemOpen
\bibfield {author} {\bibinfo {author} {\bibfnamefont {J.}~\bibnamefont
{Aalbers}} \emph {et~al.},\ }\href@noop {} {\bibfield {journal} {\bibinfo
{journal} {JCAP}\ }\textbf {\bibinfo {volume} {1611}},\ \bibinfo {pages}
{017} (\bibinfo {year} {2016})},\ \Eprint
{http://arxiv.org/abs/arXiv:1606.07001} {arXiv:1606.07001 [astro-ph]}
\BibitemShut {NoStop}%
\bibitem [{\citenamefont {Aprile}\ \emph {et~al.}(2016)\citenamefont {Aprile}
\emph {et~al.}}]{Aprile:2015uzo}%
\BibitemOpen
\bibfield {author} {\bibinfo {author} {\bibfnamefont {E.}~\bibnamefont
{Aprile}} \emph {et~al.} (\bibinfo {collaboration} {XENON}),\ }\href
{\doibase 10.1088/1475-7516/2016/04/027} {\bibfield {journal} {\bibinfo
{journal} {JCAP}\ }\textbf {\bibinfo {volume} {1604}},\ \bibinfo {pages}
{027} (\bibinfo {year} {2016})},\ \Eprint
{http://arxiv.org/abs/arXiv:1512.07501} {arXiv:1512.07501 [physics.ins-det]}
\BibitemShut {NoStop}%
\bibitem [{\citenamefont {Mount}\ \emph {et~al.}()\citenamefont {Mount} \emph
{et~al.}}]{Mount:2017qzi}%
\BibitemOpen
\bibfield {author} {\bibinfo {author} {\bibfnamefont {B.~J.}\ \bibnamefont
{Mount}} \emph {et~al.},\ }\href@noop {} {\ }\Eprint
{http://arxiv.org/abs/arXiv:1703.09144} {arXiv:1703.09144 [physics.ins-det]}
\BibitemShut {NoStop}%
\end{thebibliography}

\end{document}